\documentclass[conference]{IEEEtran}
\IEEEoverridecommandlockouts
\usepackage{cite}
\usepackage{amsmath,amssymb,amsfonts}
\usepackage{algorithmic}
\usepackage{graphicx}
\usepackage{float}
\usepackage{dblfloatfix}
\usepackage{textcomp}
\usepackage[dvipsnames]{xcolor}
\usepackage{siunitx}
\usepackage{xfrac}

\def\BibTeX{{\rm B\kern-.05em{\sc i\kern-.025em b}\kern-.08em
    T\kern-.1667em\lower.7ex\hbox{E}\kern-.125emX}}

\newcommand{\centered}[1]{\begin{tabular}{l} #1 \end{tabular}}

\usepackage{scalerel}
\usepackage{tikz}
\usetikzlibrary{svg.path}

\definecolor{orcidlogocol}{HTML}{A6CE39}
\tikzset{
	orcidlogo/.pic={
		\fill[orcidlogocol] svg{M256,128c0,70.7-57.3,128-128,128C57.3,256,0,198.7,0,128C0,57.3,57.3,0,128,0C198.7,0,256,57.3,256,128z};
		\fill[white] svg{M86.3,186.2H70.9V79.1h15.4v48.4V186.2z}
		svg{M108.9,79.1h41.6c39.6,0,57,28.3,57,53.6c0,27.5-21.5,53.6-56.8,53.6h-41.8V79.1z M124.3,172.4h24.5c34.9,0,42.9-26.5,42.9-39.7c0-21.5-13.7-39.7-43.7-39.7h-23.7V172.4z}
		svg{M88.7,56.8c0,5.5-4.5,10.1-10.1,10.1c-5.6,0-10.1-4.6-10.1-10.1c0-5.6,4.5-10.1,10.1-10.1C84.2,46.7,88.7,51.3,88.7,56.8z};
	}
}

\newcommand\orcidicon[1]{\href{https://orcid.org/#1}{\mbox{\scalerel*{
				\begin{tikzpicture}[yscale=-1,transform shape]
				\pic{orcidlogo};
				\end{tikzpicture}
			}{|}}}}

\usepackage{hyperref} 
\hypersetup{hidelinks}

\begin{document}

\title{Multimodal Exponentially Modified Gaussian Oscillators \\
\thanks{This work is funded by the Hasler Foundation under project number 22027.}
}

\author{\IEEEauthorblockN{Christopher Hahne$^{\textsuperscript{\orcidicon{0000-0003-2786-9905}}}$\,, \IEEEmembership{Member, IEEE}}
\IEEEauthorblockA{University of Bern, 
Bern, Switzerland \\
christopher.hahne@unibe.ch \\
}
}

\maketitle	

\begin{abstract}
Acoustic modeling serves audio processing tasks such as de-noising, data reconstruction, model-based testing and classification. Previous work dealt with signal parameterization of wave envelopes either by multiple Gaussian distributions or a single asymmetric Gaussian curve, which both fall short in representing super-imposed echoes sufficiently well. This study presents a three-stage Multimodal Exponentially Modified Gaussian~\mbox{(MEMG)} model with an optional oscillating term that regards captured echoes as a superposition of univariate probability distributions in the temporal domain. With this, synthetic ultrasound signals suffering from artifacts can be fully recovered, which is backed by quantitative assessment. Real data experimentation is carried out to demonstrate the classification capability of the acquired features with object reflections being detected at different points in time. The code is available at \texttt{{\color{NavyBlue}\textmd{\url{{https://github.com/hahnec/multimodal\_emg}}}}}.
\end{abstract}


\section{Introduction}
\subsection{Background, Motivation and Objective}
\begin{figure}[!t]
	\centering
	\includegraphics[width=1.0\linewidth]{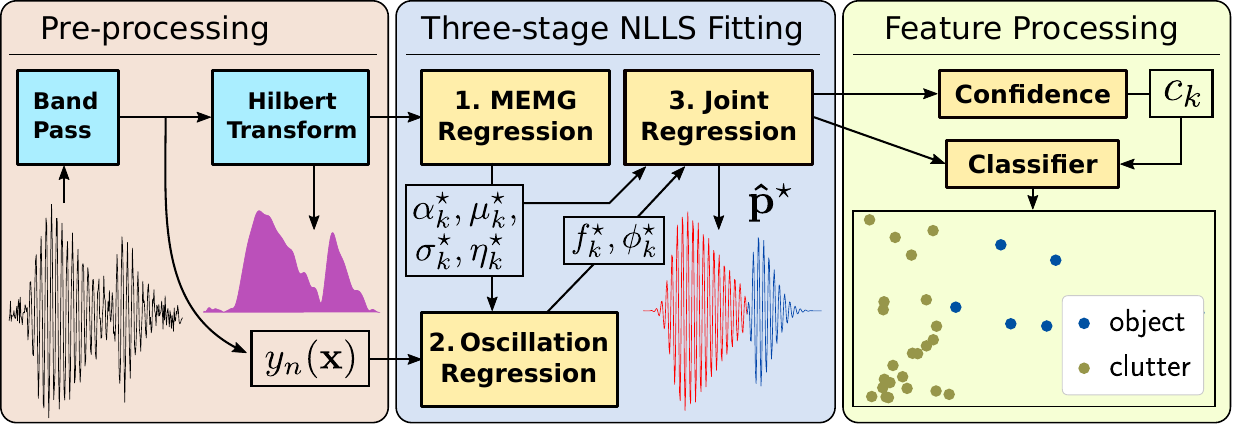}
	\caption{Overview of the proposed acoustic feature processing pipeline.\label{fig:top}}
\end{figure}

Acoustic signal modeling plays a key role in the technology stack of tomorrow’s sensor systems, with important use cases in medical analysis, non-destructive testing, robotics and vehicle navigation. 
In ultrasound sensing, real world data is characterized by a heterogeneous superposition of echoes varying in magnitude and shape as they are reflected from an environment of complex topologies. %
This work presents a robust, non-linear regression framework for acoustic feature extraction to support signal compression, denoising, data recovery, model-based testing and - most importantly - classification at low computational cost. An overview of the proposed processing pipeline is outlined in Fig.~\ref{fig:top}. \par
Echo detection in Time-of-Flight (ToF) applications is a well-studied subject across disciplines, adopting methods including dictionary-based approaches~\cite{Shi:2012,FORTINEAU:2017}, signal clusters~\cite{Mor:2015} just as Recurrent Neural Networks (RNNs)~\cite{Pan:2020}. %
Parametric sonar echo modeling is a related research subject that typically supports simulation tasks by reconstructing echoes from probability distributions. 
Pioneering work in this field was conducted by Demirli and Saniie~\cite{Demirli:1998, Demirli:2001:a}, who employed an oscillating Gaussian model to estimate single echoes by means of an iterative optimization scheme. %
A later study of the authors has taken the direction of more realistic echo representation by introducing additional parameters in an attempt to cover asymmetric envelope shapes~\cite{Demirli:2014, Demirli:2006}. %
It is worth noting that these asymmetric Gaussian regression models find application in other domains such as laser beam modeling~\cite{Li:2019}. %
%
Despite the existence of these fundamental models, echo-based feature extraction and classification remain under-investigated and have only recently attracted attention from the field of computational biology~\cite{Eliakim:2018}. %
Classification based on ultrasound signals has traditionally been addressed by means of wavelet coefficients serving as acoustic features~\cite{Meyer:1995, Tsui:2006}. %
Zhang~\textit{et al.} conducted a survey on object classification for breast cancer detection  based on two features (sound speed and attenuation) in simulated ultrasound B-scan images. The authors combined the Conjugate Gradient (CG) with a Bayesian framework and Neural Network (NN) as classifiers and concluded that the CG-NN outperforms a single feature Gauss-Newton-based classification~\cite{Zhang:2002}. %
More recently, Sterling~\textit{et al.} use a uni-modal Exponentially Modified Gaussian (EMG) at several audible frequencies to obtain object features by using an MLE optimization framework with the overall goal to make predictions on 5 different material classes~\cite{Sterling:2019}. While mainly focusing on the damping and frequency parameters in a synthetic dataset of adequate size, the authors claim to achieve a reasonable classification accuracy comparable to human perception. %
%
\par
\subsection{Statement of Contribution/Methods}
%
While these models serve as solid groundwork for model-based simulations, their performance falls short in object recognition applications such as presence detection scenarios. %
This requires realistic modeling of multiple components as materials absorb, transmit and reflect transducer signals in various ways, yielding different energy distributions that expose a multimodal representation of asymmetrically shaped Gaussian curves. %



For this purpose, this paper introduces a Multimodal Exponentially Modified Gaussian (MEMG) model with an oscillating term to regard multiple echoes as univariate probability distributions in the time domain. This acoustic feature model is sub-divided into three-stages of Non-Linear Least-Squares (NLLS) regression with each being solved by the Levenberg-Marquardt (LM) algorithm to achieve fast convergence. As the proposed framework offers characteristic features, it thereby enables echo segmentation and lays the groundwork for classification tasks.

The experimental section complements existing work by empirically demonstrating that a single skew term borrowed from stochastic calculus sufficiently represents asymmetric shapes of recorded chirp reflections. This supports the reconstruction of overlapped phase signals corrupted by noise and qualifies MEMG to be used for object classification purposes. In particular, feeding MEMG features to a Random Forest classifier shows that the proposed model helps recognize echoes across frames acquired at different points in time.
\section{Acoustic Feature Model}
\subsection{Oscillating Exponentially Modified Gaussian Components}

Let an EMG with oscillation term be defined as
\begin{align}
m(\mathbf{p}; \mathbf{x})= \alpha\,\mathcal{N}(\mathbf{x}|\mu,\sigma) \, \Phi(\mathbf{x}|\eta,\mu,\sigma) \, A(\mathbf{x}|\mu,f,\phi) \label{eq:model}
\end{align}
where $\mathbf{x} \in \mathbb{R}^{X}$ denotes the time domain with a total number of $X$ samples and $\mathbf{p}~=~\left[\alpha,\mu,\sigma,\eta, f, \phi\right]^\intercal~\in~\mathbb{R}^{D}$ contains the normalization~$\alpha$, mean~$\mu$, spread~$\sigma$, skew~$\eta$, frequency~$f$ and phase~$\phi$. With this, the Gaussian function is given by
\begin{align}
\mathcal{N}(\mathbf{x}|\mu,\sigma)=
\exp\left(-\frac{\left(\mathbf{x}-\mu\right)^2}{2\sigma^2}\right) \label{eq:gaussian}
\end{align}
and the term $\Phi(\mathbf{x}|\eta,\mu,\sigma)$ covering an asymmetric shape with
\begin{align}
\Phi(\mathbf{x}|\eta,\mu,\sigma)=\left(1+ \text{erf}\left(\eta\frac{\mathbf{x}-\mu}{\sigma\sqrt{2}}\right)\right) \label{eq:skew}
\end{align}
by means of the error function $\text{erf}(\cdot)$. The oscillating function is obtained by
\begin{align}
A(\mathbf{x}|\mu,f,\phi) = \cos\left(2 \pi f \left(\mathbf{x} - \mu\right) + \phi\right)
\end{align}
using the cosine. 
To consider superpositions of multiple components, we aggregate~EMGs by
\begin{align}
M\left(\mathbf{\hat{p}};\mathbf{x}\right)=\sum_{k=1}^K m\left(\mathbf{p}_k;\mathbf{x}\right)
\end{align}
with $\mathbf{\hat{p}}=\left[\mathbf{p}_1^\intercal, \mathbf{p}_2^\intercal, \dots, \mathbf{p}_K^\intercal\right]^\intercal~\in~\mathbb{R}^{DK}$ concatenated parameters and $k=\{1,2,\dots,K\}$ components, accordingly. \par
Prior to the MEMG model estimation, an oscillating signal is routed through a band-pass filter that eliminates spectral intensities around the dominant frequency detected by a Fourier transform. A second pre-processing stage counteracts the power loss over distance, which is caused by anisotropic radiation. Each incoming frame $y(\mathbf{x})$ is treated with an exponential fit of $\sfrac{a}{y(\mathbf{x})^b}$ where $a$ and $b$ are fitted variables.
\subsection{Three-Stage Non-Linear Least Squares Objective}
To assess candidates $\mathbf{\hat{p}}$, a loss function $L\left(\mathbf{\hat{p}}\right)$ is defined as
\begin{align}
L\left(\mathbf{\hat{p}}\right)=\left\lVert y_n(\mathbf{x})-M\left(\mathbf{\hat{p}};\mathbf{x}\right)\right\rVert_2^2 = \mathbf{f}\label{eq:minimization}
\end{align}
where $\mathbf{f}\in \mathbb{R}^{X}$ is the residual vector and $y_n(\mathbf{x})$ represents the band-pass filtered measurement data at frame number $n$. The objective in \eqref{eq:minimization} is minimized using LM steps given as
\begin{align}
\mathbf{\hat{p}}^{(j+1)} = \mathbf{\hat{p}}^{(j)} - \left(\mathbf{J}^\intercal\mathbf{J}+\delta\mathbf{D}^\intercal\mathbf{D}\right)^{-1}\mathbf{J}^\intercal\mathbf{f}
\end{align}
with the Jacobian $\mathbf{J}\in\mathbb{R}^{X\times DK}$ w.r.t. $\mathbf{\hat{p}}^{(j)}$ at each iteration $j$ and diagonal matrix $\mathbf{D}~=~\text{diag}\left[\mathbf{J}^{\intercal}\mathbf{J}\right]$ with an adaptive damping term $\delta$. Each Jacobian is computed from analytical derivatives. The optimal solution vector $\mathbf{\hat{p}}^\star$ is obtained by
\begin{align}
\mathbf{\hat{p}}^\star=\underset{\mathbf{\hat{p}}^{(j)}}{\operatorname{arg\,min}} \, L\left(\mathbf{\hat{p}}^{(j)}\right) \label{eq:goal}
\end{align}
representing the best approximation out of $\mathbf{\hat{p}}^{(j)}$ candidates. \par
Iterative optimizations are known to be error-prone for start values with large numerical distance from the solution. For a robust regression, LM iterations are split into a three-stage process with (a) Hilbert-only EMG parameter regression of $[\alpha_k,\mu_k,\sigma_k,\eta_k]$, which are used for (b) oscillation parameter optimization $[f_k,\phi_k]$ and (c) a joint parameter minimization of $\mathbf{p}_k$ at a final step. Initial positional estimates $\mu^{(1)}_k$ are obtained by using a threshold $\tau$ for the gradient of the Hilbert-transformed magnitude $\nabla_x \left|\mathcal{H}\left[y_n(\mathbf{x})\right]\right|$ as seen in
\begin{align}
\mu^{(1)}_k = \{ x_i | x_i \in \mathbf{x} \wedge \nabla_x \left|\mathcal{H}\left[y_n(\mathbf{x})\right]\right| > \tau \}
\end{align}
where $\mathcal{H}\left[\cdot\right]$ denotes the Hilbert transform. Other parameters are initialized for all $k$ with $\alpha^{(1)}_k=y(\mu^{(1)}_k)$, \mbox{$\sigma^{(1)}_k=\mathbf{1}$}, \mbox{$\eta^{(1)}_k=\mathbf{0}$}, $f^{(1)}_k=f_e$ and $\phi^{(1)}_k=\mathbf{0}$ where $f_e$ is the operating frequency given in \si{\kilo\hertz} to avoid a large condition number in the Jacobian. Phase estimates are constrained to be $\phi^{(j)}_k\in\left(-\pi, \pi\right]$.

\subsection{Feature Processing}\label{ssec:features}

Given the extracted echo feature dimensions, one may infer quantitative information about their estimation reliability.

\subsubsection{Confidence Measures}

For blind MEMG assessment, we define the confidence $C_n$ per \mbox{A-scan} by
\begin{equation}
	C_n = \left(\left\lVert\frac{M(\mathbf{\hat{p}}^\star; \mathbf{x})}{\max\left({y_n(\mathbf{x})}\right)} - \frac{y_n(\mathbf{x})}{\max\left({y_n(\mathbf{x})}\right)}\right\rVert_2\right)^{-1} \,, \quad \forall n \label{eq:frame_conf}
\end{equation}
with an inverted $\ell_2$ norm $(\lVert\cdot\rVert_2)^{-1}$ so that large values indicate higher certainty. Similarly, this can be expressed as a per-component confidence $c_k$ which writes
\begin{equation}
	c^{(n)}_k = \left(\left\lVert \hat{M}(\mathbf{p}^\star_k,\mathbf{x}) - \hat{y}_n(\mathbf{p}^\star_k, \mathbf{x}) \right\rVert_2\right)^{-1} \,, \quad \forall (k, n) \label{eq:echo_conf}
\end{equation}
where $\hat{M}(\cdot)$ and $\hat{y}_n(\cdot)$ denote the fitted and raw signals in the range of the detected position $\mu^{\star}_k$ at $5\sigma^{\star}_k$, respectively.


\subsubsection{Standardization}

Let a solution parameter vector be reshaped to a matrix $\mathbf{\hat{p}}^\star \in\mathbb{R}^{DK} \rightarrow \mathbf{\hat{P}}^{(n)}_k \in \mathbb{R}^{D\times K}$ with $K$ components and $D$ features that belongs to a set $\{\mathbf{\hat{P}}^{(1)}_k, \mathbf{\hat{P}}^{(2)}_k, \dots, \mathbf{\hat{P}}^{(N)}_k\}$ with $N$ as the total number of \mbox{A-scan} frames. After removal of components suffering from feature outliers, each $\mathbf{\hat{P}}^{(k)}_n$ is standardized according to

\begin{equation}
\mathbf{P}^{(n)}_k = \frac{\mathbf{\hat{P}}^{(n)}_k - \boldsymbol{\mu}_s}{\boldsymbol{\sigma}_s}\, , \quad \forall (k, n)
\end{equation}
with $\boldsymbol{\mu}_s\in\mathbb{R}^{D\times1}$ and $\boldsymbol{\sigma}_s\in\mathbb{R}^{D\times1}$ for valid feature evaluation.

\subsubsection{Object Detection Classifier}




The classification goal is to distinguish between pronounced echoes and reverberation clutter based on the acquired feature space. More specifically, the object detection task is formulated as an identification of unimodal components across frames whereas the classifier is unaware of the acquisition times, i.e. to which frame an echo belongs to. This implies that positional features $\mu_k^\star$ are left out in the classification to enable detection of echoes at different points in time independent of the ToF.

Using a set of standardized echo features $\mathbf{\hat{P}}^{(n)}_k$ for training, corresponding components are determined in test frames via the Random Forest (RF) classifier as an ensemble learning method. The estimation parameters, e.g. number of decision trees, are selected based on stratified cross-validation and can be found in the experimental section~\ref{sec:obj_detect}.


\section{Experimental Work}
Validation of the presented framework is conducted by simulation and experimentation. First, the model's reconstruction capability is tested by means of a simulated oscillating MEMG signal suffering from noise with quantitative analysis from the Peak-Signal-to-Noise-Ratio (PSNR). Second, real ultrasonic sensor data is used to scrutinize the relevance of the skew parameter $\eta$ introduced in \eqref{eq:model}. Finally, a set of features extracted from several real data frames is fed to proposed feature metrics to demonstrate object detection.
\subsection{Denoising from Simulation}
For denoising evaluation, a simulated Ground-Truth (GT) signal $g(\mathbf{x})=\sum_{k=1}^{K=4} m\left(\mathbf{p}_k;\mathbf{x}\right)$ with 8-bit signed integer quantization at a sampling rate $f_s=300~\si{\kilo\hertz}$ is used. 
%
%
Quantitative validation is carried out using a PSNR given by
\begin{equation}
\text{PSNR}(s(\mathbf{x})) = 20 \log_{10} \left((2^8-1)/\lVert g(\mathbf{x})-s(\mathbf{x})\rVert_2\right)
\end{equation}
where $s(\mathbf{x})$ is a measured signal. Equation~\eqref{eq:gaussian} is used for  \mbox{$g(\mathbf{x})+\mathcal{N}(\mathbf{x}|0,10)$} to mimic noisy measurements. \par
\begin{figure}[h!]
	\centering
	\includegraphics[width=\linewidth]{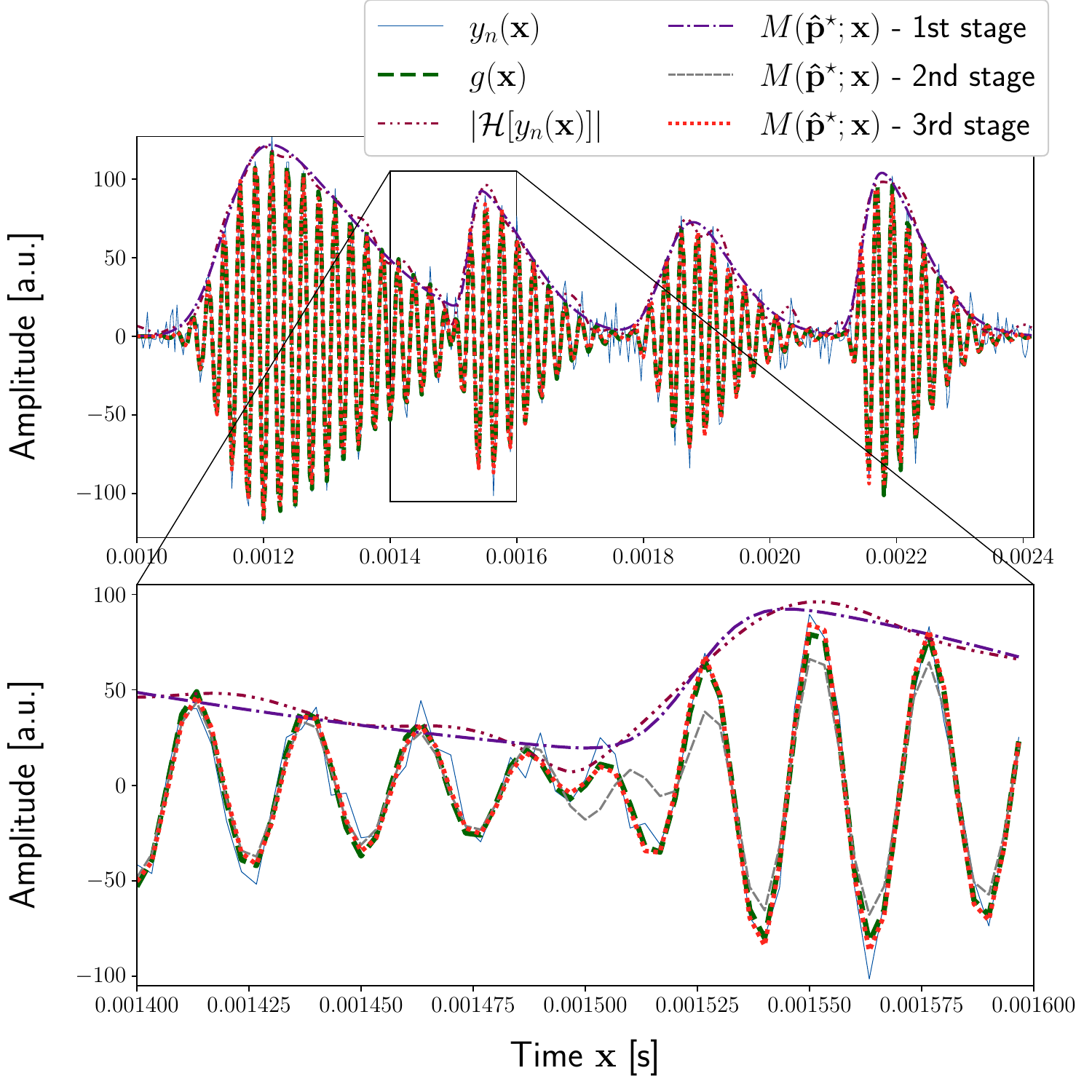}
	\caption{\textbf{Denoising} simulated chirp echoes in $y_n(\mathbf{x})$ with a $39~\si{\deci\bel}$ PSNR gain. The lower close-up indicates that recovery with model $M(\mathbf{\hat{p}}^\star;\mathbf{x})$ is even possible when echoes vary in $f$ and $\phi$ for the ground-truth signal $g(\mathbf{x})$.\label{fig:denoising}}
\end{figure}

Figure~\ref{fig:denoising} depicts the denoising result obtained by gradient sample separation of 20 and $\tau=0.1$ with a quantitative $\text{PSNR}(y(\mathbf{x}))= 65.06~\si{\deci\bel}$ for the raw data and  $\text{PSNR}(M\left(\mathbf{\hat{p}};\mathbf{x}\right))=104.91~\si{\deci\bel}$ for the oscillating MEMGs achieving a PSNR gain of $\approx40~\si{\deci\bel}$. It is believed that the presented model will perform similarly well when recovering sound waves where a portion of samples is corrupted or went missing.
\subsection{Model Performance from Real Data}
\label{sec:obj_detect}
%
%
For experimental analysis, real sensor data is acquired from a single airborne transducer at $f_e= 175~\si{\kilo\hertz}$ operating frequency, which collects \mbox{A-scan} frames of $L=126$ samples every 100~\si{\milli\second}. A convex-shaped piece of metal with \mbox{4-by-7}~\si{\centi\meter} size served as a target for a total number of $N=21$ frames. The heuristic conventions made in the evaluation are explained hereafter. Samples within the blind zone are omitted. The gradient is computed by sample distance 1 and $\tau=100$. Frequency and phase are disregarded as commonly done in NDT. LM iterations are limited to 200 each. 
Feature outliers are identified and rejected by $\mu^\star_k < 0$ and $\sigma^\star_k < 0$ while all remaining components undergo further assessment. \par
\begin{figure}[h!]
	\centering
	\includegraphics[width=\linewidth]{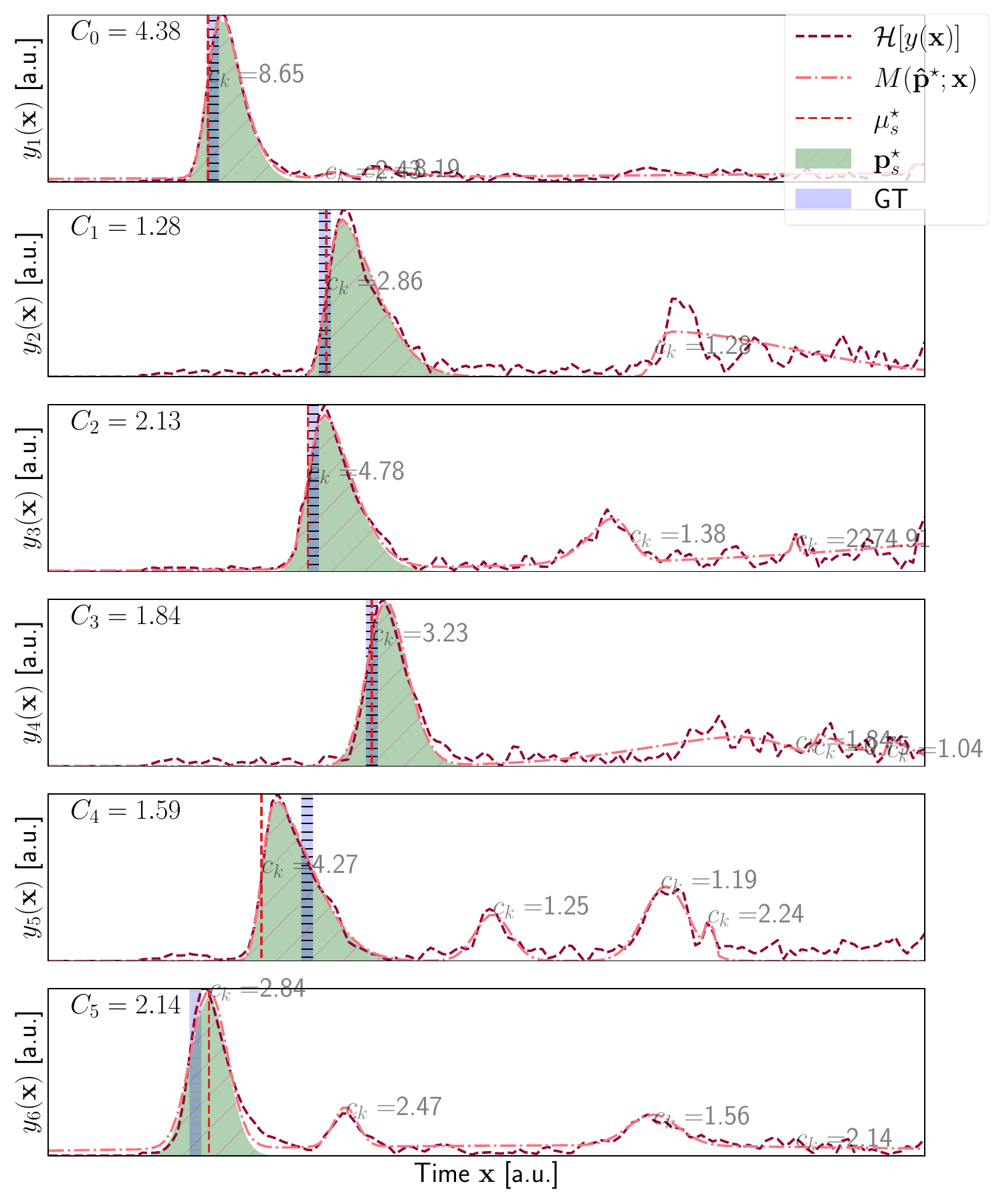}
	\caption{\textbf{Object tracking} in real data frames of the test set showing the Ground-Truth~(GT) and $\mathbf{p}^{\star}_s$ detected by a MEMG-based Random Forest classifier.\label{fig:correspondence}}
\end{figure}

One objective of the experimental validation is to scrutinize the importance of the skew parameter $\eta$. Exemplary frame confidences $C_n$ and $c^{(n)}_k$ from real acquisitions are provided in Table~\ref{tab:params}. The result suggests that skew values $\eta \neq 0$ help create a more realistic estimation when compared to previous work~\cite{Demirli:1998, Demirli:2001:a}. Moreover, this outcome backs the study by Sterling~\textit{et~al.}~\cite{Sterling:2019} who explored ring-down shapes of audible signals to distinguish between object materials. Figure~\ref{fig:correspondence} depicts frames from the test set for qualitative inspection. \par
\begin{table}[h!]
	\centering
	\caption{Confidences averaged over $N=21$ frames.}
	\resizebox{\linewidth}{!}{\begin{tabular}{c|c|c}
			 & Multimodal Gaussian~\cite{Demirli:2001:a} & \mbox{MEMG} [Ours] \\ 
			\hline
			\centered{$\frac{1}{N}\sum_{n=1}^{N} C_n$} & 1.606 & \textbf{1.804} \\
			\hline
			\centered{$\frac{1}{NK}\sum_{n=1}^{N}\sum_{k=1}^{K} c^{(n)}_k$} & 27.090 & \textbf{84.059} \\
	\end{tabular}}
	\label{tab:params}
\end{table}
\newpage
Classification of MEMG features is validated through the detection of a distinct reflector across frames. Here, $\mu_k^{\star}$ is left out to test how well features are capable of recognizing objects regardless of their distance. The power loss from anistropic radiation is compensated with $a=140.18$ and $b=1.16$. Frames are split into train and test sets by fractions of 0.7 and 0.3, respectively. Only 10 RF trees with maximum depth of 6, minimum of 1 sample per leaf and 2 per split are used.\par
\begin{figure}[h!]
    \centering
	\begin{minipage}{.41\linewidth}
	\centering
	\includegraphics[width=\linewidth]{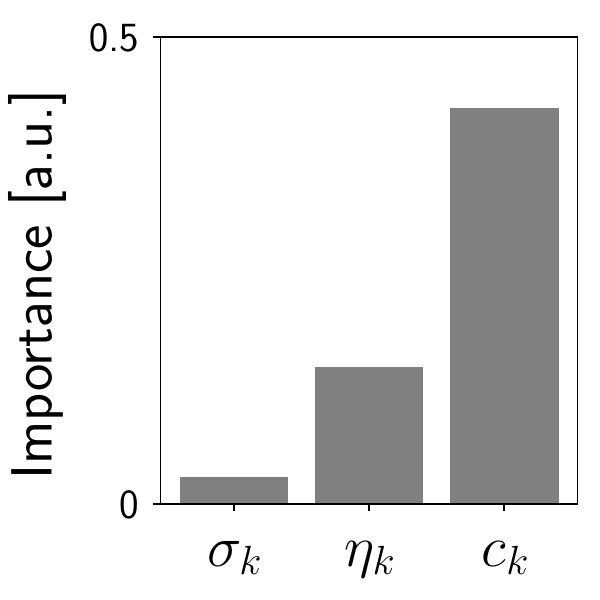}
	\label{fig:feat_importance}
	\end{minipage}	
	\begin{minipage}{.41\linewidth}
		\centering
		\includegraphics[width=\linewidth]{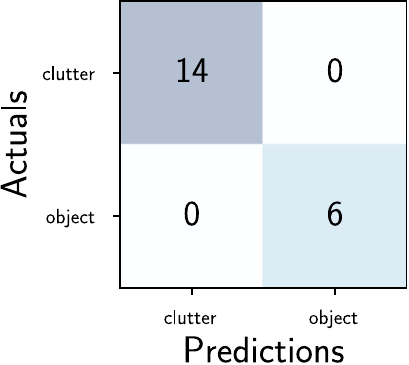}
		\label{fig:conf_mat}
	\end{minipage}%
	\caption{\textbf{Classification} of objects from Fig.~\ref{fig:correspondence} with feature importance (left) and confusion matrix (right) in absence of any $\alpha_k$ and $\mu_k$.\label{fig:interpret_class}}
\end{figure}

%
Figure~\ref{fig:correspondence} shows qualitative results where the GT from the transducer and detected echo components $\mathbf{p}^{\star}_s$ are highlighted. 
Prediction is assessed by the F-score, which yields a \mbox{$F_1 = 1.0$} when feeding $\alpha^{\star}_k$, $\sigma^{\star}_k$, $\eta^{\star}_k$, $c_k$. The impact of $\alpha$ is further analyzed by entirely excluding it from the RF classifier still giving \mbox{$F_1 = 1.0$}. A confusion matrix and feature importance diagram are shown in Fig.~\ref{fig:interpret_class} for intuitive interpretability of the class prediction. 
\section{Conclusions}

This study shows that super-imposed ultrasound echoes are robustly represented by an accumulation of oscillating EMGs. Quantitative assessments confirm this hypothesis in a simulation analysis by reconstructing a signal with a $40~\si{\deci\bel}$ PSNR gain. Real data experimentation is carried out to demonstrate the relevance of Gaussian skew. The echo classification capability is supported by an F1-score of 1.0 regardless of the arrival time. This suggests that the presented framework is a viable acoustic feature extractor in ToF applications. \par



\bibliographystyle{IEEEtran}
\bibliography{SPaCE_literature}

\end{document}